\documentstyle[aps,prl,epsfig]{revtex}
\begin{document}
\twocolumn[\hsize\textwidth\columnwidth\hsize\csname @twocolumnfalse\endcsname

\title{ The role of the dopant in the superconductivity of diamond }

\author{X. Blase, Ch. Adessi, D. Conn\'etable}

\address{ Laboratoire de Physique de la Mati\`ere Condens\'ee et des
Nanostructures (LPMCN), CNRS and Université Claude Bernard Lyon I, UMR 5586,
   Bâtiment Brillouin, 43 Bd du 11 Novembre 1918, 69622 Villeurbanne Cedex, France.}

\date{\today}
\maketitle

\begin{abstract}
We present an {\it ab initio} study of the recently discovered superconductivity 
of boron doped diamond within the framework of a phonon-mediated pairing mechanism.
The role of the dopant, in substitutional position, is unconventional in that
half of the coupling parameter $\lambda$ originates in strongly localized
defect-related vibrational modes, yielding
a very peaked Eliashberg $\alpha^2F(\omega)$ function. 
The electron-phonon coupling potential is found to be extremely large
and T$_C$ is limited by the low value of the density of states at the Fermi level.
\end{abstract}



\narrowtext
\vskip2pc]


In a recent paper~\cite{Nature04}, the superconductivity of boron doped (B-doped)
diamond was evidenced experimentally with a transition temperature
T$_C$ of 4~K. This work follows the discovery of
superconductivity in doped silicon 
clathrates ~\cite{kawajiPRL95,conneta03} (T$_C$=8~K),  a cage-like silicon
material which crystallizes in the same {\it sp}$^3$ environment
as the diamond phase. Further, the superconductivity of carbon-based
clathrates was predicted on the basis of {\it ab initio} simulations
\cite{conneta03,BenedekSupra}.

Even though the reported temperatures are rather low,
the superconducting transition of column IV semiconductors
is of much interest as it concerns very common materials.
Further, in the case of the carbon clathrates, it has been predicted
that the electron-phonon potential V$_{ep}$=$\lambda$/N(E$_F$)
(where N(E$_F$) is the density of states at the Fermi level), 
may be extremely large \cite{conneta03,BenedekSupra}.  
Finally, the superconductivity of hole-doped strongly bonded
covalent systems is reminiscent of that of the MgB$_2$ compound
~\cite{MgB2}.

As compared to the cage-like clathrate phase, diamond is difficult to
dope as the network is much denser. This explains why the superconductivity
of doped diamond was never observed, as only quite recently
a doping in the limit of a few percent could be achieved. As described in
Ref.~\onlinecite{Nature04}, it is only for a boron concentration c $>$ 10$^{22}$
cm$^{-3}$ that a degenerate semiconductor or metallic character can be observed.

In two recent work~\cite{Boeri,Pickett}, a first theoretical  study of the
superconductivity of B-doped diamond within an elegant virtual crystal
approximation was proposed.
In this approach, the crystal is made of ``averaged carbon-bore atoms" (nuclear
charge Z=(1-x)Z$_C$ + xZ$_B$) within the standard 2 atoms/cell structure of diamond,
and B atoms are not considered explicitly. As such, the superconductivity
of B-doped diamond was analyzed in terms of Bloch-like delocalized electron
and phonon ``average" states.

In this work, we study by means of {\it ab initio} calculations the
electron-phonon coupling in B-doped diamond. We consider explicitly 
a large  supercell with one B atom in substitution. The metallic
character of the doped system is evidenced and the deviations from
a simple rigid band model are discussed.
We show further that the boron atoms yield very localized vibrational modes
which contribute to half of the electron-phonon coupling parameter $\lambda$,
in large contrast
with the picture conveyed by the virtual crystal approximation.  
The value of $\lambda$ is calculated {\it ab initio},
yielding an unusually large electron-phonon coupling potential V$_{ep}$
or deformation potential D.

Our calculations are performed within the local density approximation
to density functional theory~\cite{hohenbergPRB64} (DFT)
and a pseudo-potential~\cite{troullierPRB91} plane-wave approach.
A 55~Ryd energy cut-off and a (4x4x4) Monkhorst-Pack~\cite{monkhorstPRB76}
sampling of
the Brillouin zone (BZ) showed good convergence for structural
relaxations. The  phonon modes  and the electron-phonon (e-ph) coupling
matrix elements were obtained within the framework of 
perturbative DFT~\cite{baroni}. Due to the computational cost,
phonons were calculated at the $\bf q=\Gamma$ point only.
This is equivalent to a (3x3x3) sampling of the undoped diamond
BZ.

For the calculation of the average e-ph matrix elements $g_{q\nu}$ over
the Fermi surface, required to evaluate $\lambda$:

\begin{eqnarray}
\lambda =  N(E_f) V_{ep} 
        = 2 N(E_f) \sum_{{\bf q}\nu} < |g_{q\nu}|^2> / \hbar \omega_{q\nu}
\\
< |g_{q\nu}|^2> = \int { d^3k \over \Omega_{BZ}}
 \sum_{n,n'} |g_{q\nu} (knn')|^2 
 { \delta(\epsilon_{k+q,n'}) \delta(\epsilon_{kn})
   \over  N(E_F)^2 }
\\
g_{q\nu}(knn') = ({\hbar \over 2M\omega_{q\nu}})^{1/2}
 <  \psi^0_{n{\bf k}} | {\bf \hat \epsilon_{q\nu}} \cdot
 {\delta V \over \delta \hat{u}_{q\nu} } | \psi^0_{n'{\bf k+q}}> 
\end{eqnarray}

\noindent
a much larger (8$\times$8$\times$8) $\bf k$-point sampling was used.
In Eqn. 2, 
the origin of the electronic energies ($\epsilon_{kn}$) is taken to be the Fermi
energy E$_F$.

Our system consists of a diamond (3x3x3) supercell (54 atoms) with one B
atom in substitution (noted BC$_{53}$ in what follows). 
We are thus in a 1.85$\%$ doping limit, close to
the 2.5 $\pm$0.5$\%$ studied experimentally.
Upon structural relaxation, the B-C bond length is 1.57 $\AA$ as compared
to the 1.54 $\AA$ for the C-C bonds in undoped diamond (LDA values).
The distance between second and third carbon neighbors is reduced to
1.52 $\AA$ in order to accommodate the larger B-C bonds.

\begin{figure}[!hbtp]
\begin{center}
\includegraphics[width=9cm,clip=]{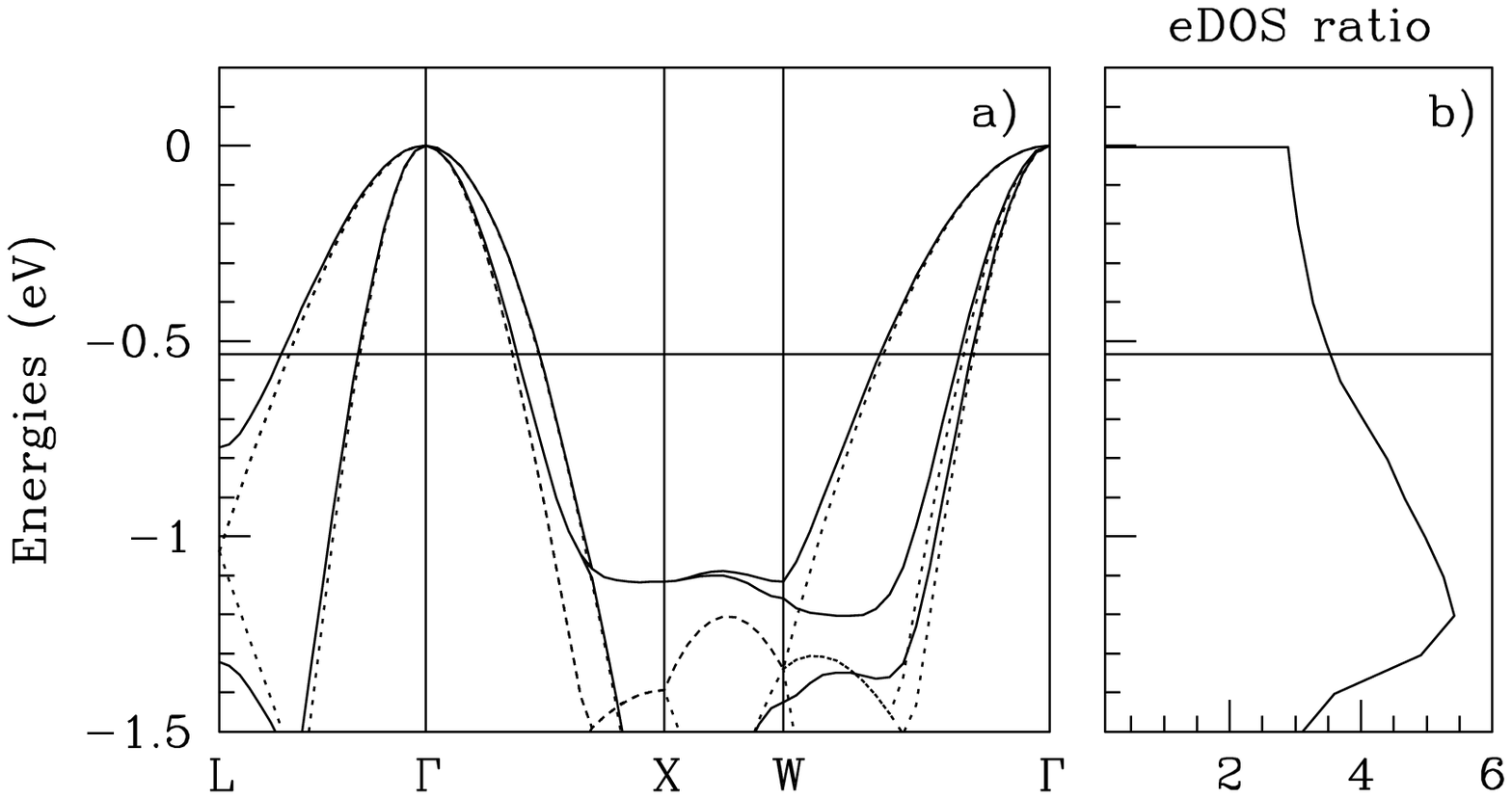}
\caption{
(a) Details of the top of the valence bands
along high-symmetry directions of the FCC BZ
for undoped diamond in the (3x3x3) cell (dotted lines) and the doped
BC$_{53}$ system (full lines). The horizontal line is the position of the Fermi level
in the doped case. The two band structures have been aligned at the top of 
the valence bands.
(b) Ratio of the density of states projected onto the B-orbitals to the total density
of states per atom in the unit cell 
(this ratio has been set to zero in the band gap).
        }
\label{fig1}
\end{center}
\end{figure}

We first explore the electronic properties and compare the band structure of
BC$_{53}$ with that of diamond (see Fig.~\ref{fig1}). As expected,
we are clearly in a metallic regime with the Fermi level located at
$\sim$ 0.54 eV
below the top of the valence bands. Even though Fig.~\ref{fig1} seems
to confirm the model of rigid band doping around  E$_F$, clear differences
exist in particular at the zone boundaries where degeneracies are left due to
the reduction of the periodicity under doping.
We note that in most directions, the band structure of
pure and B-doped diamond start to significantly depart from each other
precisely below E$_F$~\cite{BC3}.

We also calculate the total density of electronic states (eDOS) projected
onto the atomic orbitals. Even though the eDOS projected onto
the B orbitals (eDOS$_B$) is small,
it is instructive to plot (Fig. 1b) the ratio N$_a$$\times$eDOS$_B$/eDOS, 
where N$_a$=54 is the number of atoms per unit cell. Clearly, close to
E$_F$, the relative weight of B orbitals into the wavefunctions is
much larger than the weight on any given carbon atom.
As expected,  this ratio increases where
the bands of the ideal and doped diamond differ most significantly.
An analysis of the projection of the eigenstates onto the atomic orbital basis
shows that the leading coefficients are the ({\it p}$_x$,{\it p}$_y$,{\it p}$_z$)
B orbitals~\cite{spilling}. Such results indicate that even though
degenerate with the diamond Bloch states, the wavefunctions of interest
for e-ph coupling
exhibit some degree of localization around the B atom.

We now study the phonon states at $\Gamma$. The related DOS ({\it p}DOS)
is shown in Fig. 2b and compared in Fig. 2a to the one of undoped diamond
~\cite{sampling}.  As compared to the virtual-crystal approximation results,
we do not observe the very large softening of the zone-center phonon
modes predicted in Ref.~\onlinecite{Boeri}. However, the splitting of the
highest diamond mode (Fig. 2a) clearly results in
transferring spectral weight to lower energy. This can be interpreted
on the average as a softening of the optical frequencies.
In particular,
zone-center modes emerge around 1200 cm$^-1$ (Fig. 2b) which are consistent
with the appearance of new broad peaks in Raman experiments around 1220
cm$^{-1}$ \cite{Nature04,Raman}.

More insight can be gained by calculating the {\it p}DOS 
projected onto the B atom and its 4 C neighbors (thick line).
The most salient feature is that the new modes (3-fold) at 1024 cm$^{-1}$
(vertical dotted lines in Fig. 2) project to more than 50$\%$ on these 5 atoms.
The analysis of the associated eigenvectors clearly
indicates that these three modes correspond to the stretching of the B to
neighboring C atom bonds. The displacement of the B atom
is $\sim$ 22$\%$ larger than that of C neighbors as expected
from the difference in masses. Therefore, the dopant atoms induce  vibrational
modes which are very localized in space. This is a significant deviation
from a virtual crystal approach where only extended phonon states 
can be evidenced. The influence of such modes on the e-ph coupling
is now shown to be crucial.

We follow equations 1-3 to calculate $\lambda$.
With our limited $\bf q$-point sampling~\cite{gamma}, we find
that $\lambda(q=\Gamma)$ is rather sensitive to the gaussian  broadening 
used to mimic the $\delta$-functions in Eq. 2.
However, such a variation is primarily related to the variation
of the nesting factor n(q) (that can be obtained by setting  $g_{q\nu}(knn')$
to unity in Eq. 2). This factor n(q) is a measure of phase space
availability for electron scattering by phonons. 
Indeed, one verifies that the quantity $\lambda(q)/n(q)$ (q=$\Gamma$)
is much more stable and displays a plateau over a large energy broadening
range. We thus define an averaged $\lambda$ by rewriting:

\begin{eqnarray*}
 \lambda & = & \int d^3q \lambda(q) = \int d^3q \: n(q) \: \lambda(q)/n(q)  \\ 
         & \sim & \lambda(\Gamma)/n(\Gamma) \int d^3q \: n(q) 
   = N(E_F)^2 \lambda(\Gamma)/n(\Gamma)
\end{eqnarray*}

\noindent
Such an approximation relies on the usual assumption that the e-ph matrix 
elements are varying much more smoothly as a function of $\bf q$ than the 
nesting factor.

We find $\lambda$=0.43 $\pm$0.01. Such a value of $\lambda$ is much smaller
than that of MgB$_2$ (in the absence of anharmonic effects~\cite{anharmonic},
$\lambda$(MgB$_2$) $\sim$ 1). Assuming now that the McMillan 
expansion~\cite{McMillan} is valid in the limit of degenerate semiconductors:

$$
T_c = { \hbar \omega_{log} \over 1.2 k_B}
exp \left[ { -1.04 (1+\lambda) \over \lambda - \mu^* ( 1 + 0.62 \lambda) } \right]
$$

\noindent and with a logarithmic averaged phonon 
frequency $\omega_{log}$ of 1023 cm$^{-1}$,
the screening parameter $\mu^*$ needed to reproduce the 4~K
experimental value of T$_C$  is found to be of the order of 0.13-0.14
\cite{boerimcmillan}. This is a very standard value, indicating
that phonon-mediated pairing mechanisms can adequately account for the
superconducting transition in impurity-doped diamond.

With N(E$_F$)=20.8 states/spin/Ry/Cell, the e-ph potential V$_{ep}$
is 280 meV $\pm$ 10 meV.
Such a value is comparable to what was found in the Li-doped 
carbon clathrates (V$_{ep}$ $\sim$ 250 meV \cite{BenedekSupra})
and much larger than the 60-70 meV found for the fullerides
~\cite{c60ref}. 
This was analyzed in terms of the amount of {\it sp}$^3$ character
of the C-C bonds~\cite{conneta03}.
For sake of comparison with Refs.~\onlinecite{Boeri,Pickett}, we calculate the
deformation potential D such that $\lambda$=N$_2$D$^2$/M$\omega^2$, where
N$_2$ is the eDOS in eV/state/spin/``2 atoms cell".
With N$_2$=0.057 and $\omega$=1023 cm$^{-1}$,
one finds D = 27 eV/$\AA^{-1}$, larger than the D $\sim$ 21 eV/$\AA^{-1}$ value found
in the virtual crystal approximation~\cite{Boeri,Pickett}.  This value is more than
twice as large as the one calculated for MgB$_2$, indicating that it is primarily
the low density of states at the Fermi level, related to the 3D nature of diamond,
that is responsible for the low value of $\lambda$ and T$_c$.

Better $\bf q$-point sampling and the influence of anharmonicity
\cite{Boeri,anharmonic} will certainly change the value of $\lambda$.
Beyond the numerical estimates,
it is important to look at the spectral decomposition of
$\lambda$, that is the Eliashberg function $\alpha^2 F(\omega)$ (Fig. 2c):

\begin{equation}
\alpha^2 F(\omega) = {1 \over 2} \sum_{{\bf q}\nu} \omega_{{\bf q}\nu} \lambda_{{\bf q}\nu}
   \delta(\omega - \omega_{{\bf q}\nu})
\end{equation}

\noindent
By comparison with the {\it p}DOS, it appears that a significant contribution to
$\lambda$ comes from the B-related phonon modes.
{\it  As a matter of fact, these 3 modes
contribute to about 50$\%$ of the coupling}~\cite{note2}.  
In particular,
the resulting Eliashberg function is very peaked at the 
impurity vibrational modes energy, displaying a $\delta$-like shape.
Therefore, not only the density
of states N(E$_F$), but also the strength of the e-ph coupling, can be
taylored by changing the dopant concentration and/or chemical type.
This mechanism offers thus an additional flexibility in trying to
increase T$_C$. 
The situation is very different from the one of fullerides or clathrates 
where the low frequency phonon modes associated with the dopant
hardly contribute to $\lambda$.

To strengthen that point,  we  look at the superconductivity
of diamond within a rigid band model~\cite{diamondsample}. 
We plot in Fig. 3 the evolution
of $\lambda$ as a function of the position of the Fermi level. For E$_f$ located
$\sim$ 0.5-0.6 eV below the top of the valence bands, 
we find a $\lambda$ of $\sim$ 0.24,
that is roughly 55$\%$ of what we find in the BC$_{53}$ case.
Even though very qualitative,
this computer experiment substantiates the conclusion
that B-related modes contribute significantly to the e-ph coupling.
Further, Fig. 3 clearly suggests that larger $\lambda$ could be achieved 
under larger doping conditions due to the quadratic rise of N(E$_F$)
as a function of E$_F$.

It is certainly accidental that just half of the coupling originates in B-modes as the
weight of the eigenstates on the dopant should scale as $\sqrt N_a$ under varying
concentration  (neglecting the electronic localization effect described above). 
In addition, the simple structural model studied here
does not account for the effect of randomness,
nor the possibility for B atoms to cluster or adopt other configurations. 
It is clear however that the role of the dopant
cannot be interpreted solely as providing charges to the network,
nor as softening the phonon
modes by depleting the bonding states at the top of the valence bands.
In particular, such results reveal that any
attempt to explain the superconductivity of doped diamond within a rigid band 
model or the virtual-crystal approximation should be taken with care.
The importance of the impurity related vibrational modes is expected to
be very general and should apply to any semiconductor/insulator doped
by strongly bound atoms.

In conclusion, we have shown that the role of boron in doped diamond is unconventional
in that a significant fraction of the coupling coefficient $\lambda$ originates in
vibrational modes very localized on the defect centers. With a $\lambda$ of 0.43,
much smaller than that of MgB$_2$ $\sim 1$, T$_C$ remains small. Even though the
electron-phonon potential is extremely large, the 3D nature of the network 
reduces the density of states at the Fermi level. This invites to study
the case of doped diamond surfaces where both the contraction of the reconstructed 
bonds and the 2D nature of the surface states may lead to much larger T$_C$ \cite{surface}.

\noindent {\bf Acknowledgements:}
Calculations have been performed at the French CNRS national computer center
at IDRIS (Orsay). The authors are indebted to r\'egion Rh\^one-Alpes for
financial support  through the ``Th\'ematique prioritaire programme 2003-2005"
in Material Sciences.



\begin{figure}[!hbtp]
\begin{center}
\includegraphics[width=9cm,clip=]{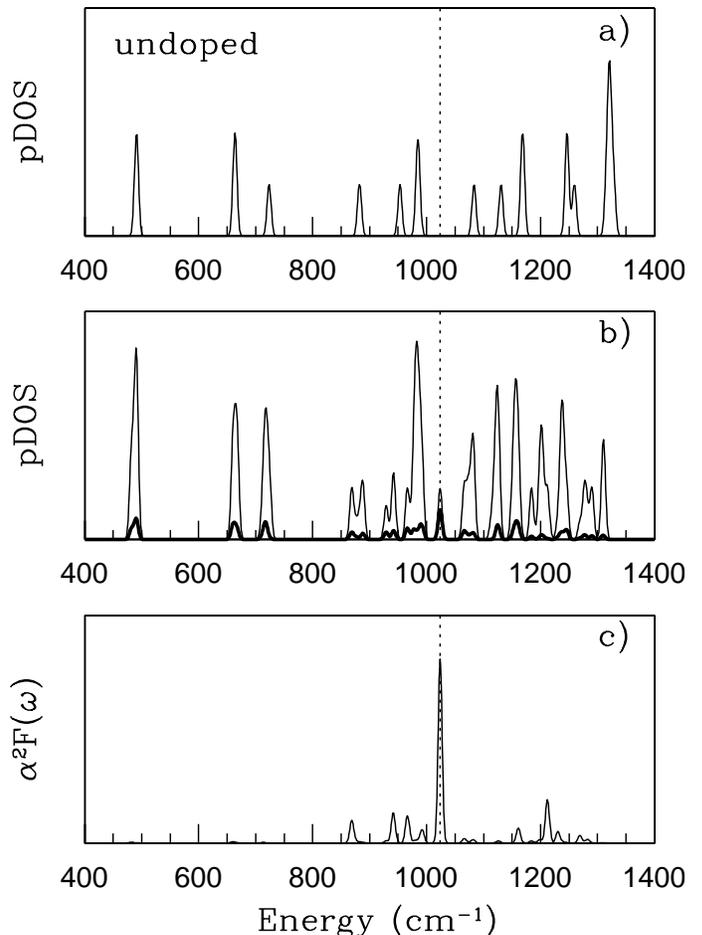}
\caption{
Phonon density of states {\it p}DOS for (a) bare diamond with
modes on an unshifted (3x3x3) grid (b) the boron
doped BC$_{53}$ cell with modes at $\Gamma$. In (c)
the spectral function $\alpha^2F(\omega)$ is represented
for modes at $\Gamma$. Energies are in cm$^{-1}$ and
the y-axis magnitude is arbitrary. A 5 cm$^{-1}$ broadening
has been used. The vertical dotted
lines indicate the energy position of the 3-fold B-related
vibrational modes.
        }
\label{fig2}
\end{center}
\end{figure}

\begin{figure}[!hbtp]
\begin{center}
\includegraphics[width=9cm,clip=]{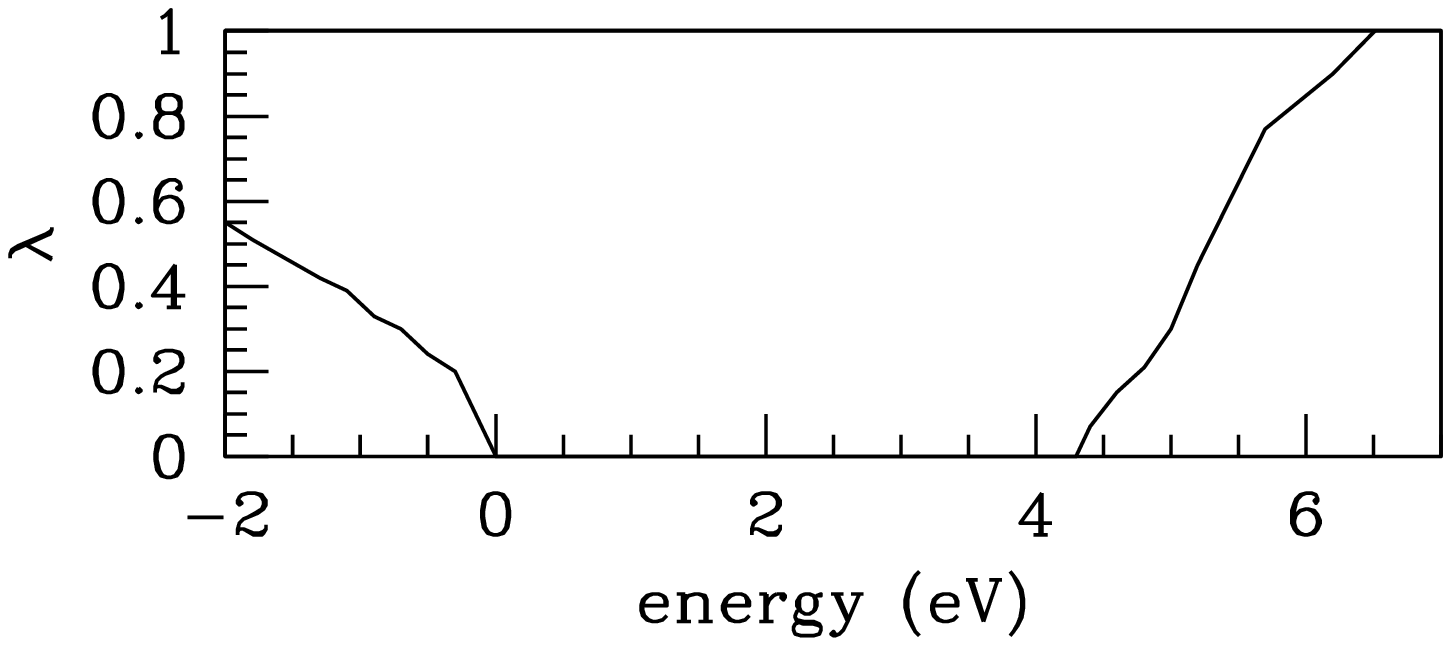}
\caption{
Evolution of $\lambda$ for diamond, doped within a rigid band
model (see text), as a function of the Fermi energy position E$_F$.
The energy reference 
has been set to the top of the valence bands.
        }
\label{fig3}
\end{center}
\end{figure}


\begin{references}
\bibitem{sampling}
For undoped diamond, we use the conventional 2 atoms/cell structure with
a (3x3x3) $\bf q$-grid which folds onto the $\Gamma$ point of the
(3x3x3) cell.
\bibitem{gamma}
In the limit of low doping, and as emphasized in Refs.~\onlinecite{Boeri,Pickett},
zone-center
optical phonons are expected to contribute primarily to the e-ph coupling. 
Such an approximation is at the heart of the q=$\Gamma$ deformation potential
approach developed by Lee and Pickett in Ref.~\onlinecite{Pickett}.
\bibitem{boerimcmillan}
Using the modified McMillan formula proposed in Ref.~\onlinecite{Boeri} for $\delta$-shaped
Eliashberg functions (see Fig. 2c), we find a $\mu^*$ of 0.13 for $\omega$=1023 cm$^{-1}$
in good agreement with the standard formulation used in the text.
\bibitem{note2}
This explains in particular that $\omega_{log}$ is found to be close to these
modes energy.
\end{references}
\end{document}